# Revisiting Kepler's Laws of Equal Areas and Ellipses for the Earth

Wu-Yi Hsiang[a], Hai-Chau Chang[b], Herng Yao[c,]*, Pon-Jen Chen[c]

[a]Department of Mathematics, University of California, Berkeley, CA 94720, USA

[b]Department of Mathematics, National Taiwan University, Taipei, Taiwan 106, ROC

[c]Department of Physics, National Taiwan Normal University, Taipei, Taiwan 116, ROC

**Abstract**

Kepler's laws of planetary motion are acknowledged as highly significant to the construction of universal gravitation. The present study demonstrates different ways to derive the law of equal areas for the Earth by general geometrical and trigonometric methods, which are much simpler than the original derivation depicted by Kepler. The established law of equal area for the Earth was applied to analyze the angular velocity —or the reciprocal of the distance—for the Earth's orbit around the Sun, and can be defined as a periodic function by analyzing the available data, which helps explain the law of ellipses for the Earth.

## 1. Introduction

It has been 400 years since Johannes Kepler (1571–1630) published his major masterpiece — *New Astronomy* [1]. That book proposed a concise description of the orbit of Mars. Mars orbits the Sun in an ellipse with the Sun located at one focus rather than in accordance with the long-held belief that Mars moved in a perfect circle. Also, the line connecting the Sun and Mars sweeps an equal area in an equal period of time. These discoveries are known as Kepler's first and second laws of planetary motion, or the laws of ellipses and equal areas, respectively. These findings, of course, depended on the legacy of precise observations by Tycho Brahe (1546–1601). More importantly, it was Kepler's creativity that led him to apply geometric methods to convert astronomical data based on an Earth-centered model to a Sun-centered model.

The first part of this article is intended to demonstrate a geometrical method, which is much simpler than the original depicted by Kepler, to obtain the law of equal areas using mainly the law of sines that every high school student learns. In the second part of this article, the established law of equal areas for the Earth is applied to calculate the angular velocity of the Earth around the Sun. It can be shown to be a periodic function by analyzing the data using current mathematical methods, and this characteristic may assist in understanding the law of ellipses for the Earth.

## 2. The Law of Equal Areas for the Earth

To construct the law of equal areas for the Earth, we must first understand the position of the Earth relative to the Sun. However, how do we determine the Earth's location in the Universe? From thousands of years ago until the era of Kepler, all of the understanding about the motions of the planets and the Sun was based on observations taking the Earth as the origin from which to measure the angles or positions of the planets and the Sun. Specifically, man could only record the positions of the planets and the Sun as longitudes and latitudes, but was unable to measure their distances from the Earth or calculate corresponding distance ratios. For this reason, Kepler, as a believer in the heliocentric model, needed to convert observational data based on the Earth as the origin to data based on the Sun as the origin, which required that he come up with a new approach. In Kepler's era, it was known that planets moved in the same plane. To locate an object on a plane, one must have at least two reference points as a basis. However, if the Sun is taken to be one fixed point, what can be used as a second reference point? If a fixed star is chosen, its distance cannot be determined; but if a nearby planet is used, its location in the sky is not stationary.

Nevertheless, Kepler noted that the time interval from the one opposition — where the Sun, the Earth, and Mars are aligned — to the next opposition was about 780 days. Based on this data, he then determined the period of Mars' orbit around the Sun to be approximately 687 days. Because the position of Mars in the sky repeats every 687 days, this position can be used as a reference point. As a result, there are 2 fixed points to use as a benchmark. Finally, by applying the data of angular positions observed from the Earth, one can calculate the location of the Earth relative to the Sun.

The position is denoted by its ecliptic longitude where the longitude of the spring equinox is 0°, that of the summer solstice is 90°, and so on. Kepler then chose the date of the opposition of Mars as a basic reference point. For convenience, one may select 5 a.m. on March 25, 1950, when Mars was in opposition, i.e., the Sun ($S$), Earth ($E$) and Mars ($M$) were in a straight line as shown in Fig. 1. $M$ will return to its original position every Martian year, and this can be treated as a second fixed point. $E_i$ and $E_j$ represent the positions of the Earth one Martian year before and after the opposition of Mars, respectively. From the observational data, these configurations occurred at May 5, 1948 and Feb. 8, 1952, respectively. As a result, $SE_iME_j$ forms a quadrilateral. If $r_i$ and $r_j$ are the lengths of the segments $SE_i$ and $SE_j$, then $r_i$ and $r_j$ represent distances from the Sun to the Earth at two different times (Fig. 2).

The quadrilateral $SE_iME_j$ can be treated as being made of $\triangle SE_iM$ and $\triangle SE_jM$, where $\angle SE_iM$, $\angle SE_jM$, $\angle E_iMS$ and $\angle E_jMS$ can be observed as the longitudes of the Sun and Mars as seen from the Earth. The longitudes of the Sun and Mars were 44.7° and 144.9°, respectively, as seen from the Earth, $E_i$ [2]. This implies the following equation:

$$\angle SE_iM = \mu_i = 144.9° - 44.7° = 100.2°.$$

Similarly, the longitudes of the Sun and Mars as seen from the Earth, $E$, were 218.2° and 318.3°, respectively, as seen from the Earth, $E_j$ [2]. This implies the following equation:

$$\angle SE_jM = \mu_j = 318.3° - 218.2° = 100.1°.$$

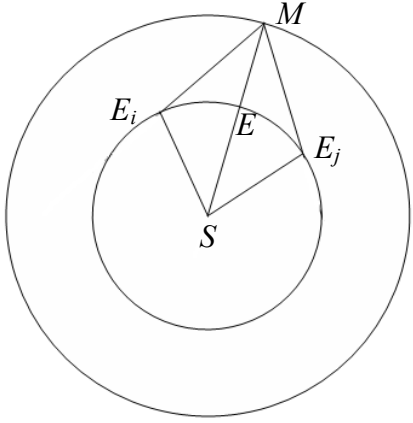 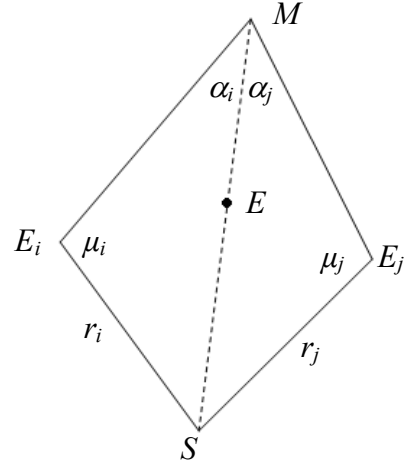

Fig. 1: The positions of the Sun(*S*), the Earth(*E*) and Mars(*M*). $E_i$ and $E_j$ represent the positions of the Earth one Martian year before and after the opposition of Mars, respectively.

Fig. 2: The quadrilateral formed by $SE_iME_j$. The angles $\mu_i, \mu_j, \alpha_i$ and $\alpha_j$ are all observed.

However, the longitude of Mars at the opposition of March 25, 1950, was 182.0°, and the longitude of Mars as seen from the Earth, $E_i$, on May 5, 1948, was 144.9°. Hence,

$$\angle E_i MS = \alpha_i = 182.0° - 144.9° = 37.1°$$

Similarly, the longitude of Mars as seen from the Earth, $E_j$, on Feb. 8, 1954, was 218.2°.

$$\angle E_j MS = \alpha_j = 218.2° - 182.0° = 36.2°$$

The 3 interior angles in $\triangle SE_iM$ and $\triangle SE_jM$ are then all known, and $SM$ is a common side. Therefore, by the law of sines,

$$\frac{r_i}{\sin \alpha_i} = \frac{SM}{\sin \mu_i}, \qquad \frac{r_j}{\sin \alpha_j} = \frac{SM}{\sin \mu_j}.$$

The relationship between $r_i$ and $r_j$ is as follows:

$$\frac{r_j}{r_i} = \frac{\sin \mu_i}{\sin \mu_j} \frac{\sin \alpha_j}{\sin \alpha_i}.$$

Originally, it was very hard to measure the actual distance between the Sun and the Earth. Now their ratio can be obtained from the corresponding angles $\angle SE_iM$ and $\angle SE_jM$ spanned by the lines connecting the Sun and Mars to the Earth 1 Martian year before and after the date of the opposition of Mars, respectively, and from the angles $\angle E_iMS$ and $\angle E_jMS$ spanned by the line of opposition of Mars and by the line connecting the Earth to Mars 1 Martian year before and after the date of the opposition of Mars, respectively. If $\omega_i$ and $\omega_j$ are the angular velocity of the Earth at $E_i$ and $E_j$ relative to the Sun, they can be calculated by the angles swept by the Earth in one day after the dates at $E_i$ and $E_j$, respectively. Since the longitudes of the Earth as seen from the Sun on May 5, 1948 and May 6, 1948 are 224.897° and 225.865°, respectively, the angular speed of $\omega_i$=225.866–224.897=0.969. Similarly, $\omega_j$=139.561–138.549=1.012.

The law of equal areas for the Earth means the line connecting the Earth to the Sun sweeps an equal area in the same period of time, as shown in Fig. 3. Namely,

$$\Delta A = \frac{1}{2} r^2 \Delta\theta$$

and

$$\frac{\Delta A}{\Delta t} = \frac{1}{2} r^2 \frac{\Delta\theta}{\Delta t}.$$

So the areal velocity is

$$\frac{dA}{dt} = \frac{1}{2} r^2 \frac{d\theta}{dt} = \frac{1}{2} r^2 \omega,$$

where $\omega$ is the angular speed of the Earth around the Sun.

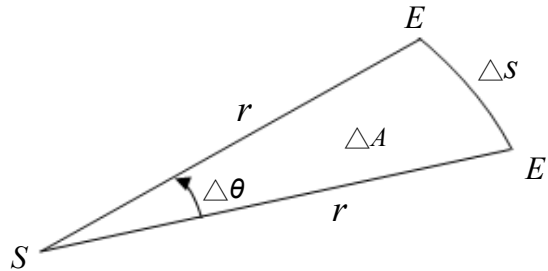

Fig. 3: The law of equal areas for the Earth for a short period of time.
The distance from the Earth to the Sun can be treated as a constant.

Hence, in order to prove the law of equal areas it is only necessary to show that the product of the square of the distance from the Earth to the Sun and the corresponding angular speed of the Earth is a constant, i.e.,

$$r_i^2 \omega_i = r_j^2 \omega_j \qquad (1)$$

The position of the Earth $E_i$, which occurred 1 Martian year before the opposition of Mars as a reference point, is selected. Then 6 more positions of the Earth, $E_j$, are chosen, with 3 of them occurring 1, 2 and 3 Martian years before the time of the reference point, $E_i$. From observations of these $E_j$'s, one can find the corresponding values of the ratios $r_j^2/r_i^2$ and $\omega_i/\omega_j$, as shown in Table 1.

From the calculating results in the right 2 columns in the table, one may see that the difference between $r_j^2/r_i^2$ and $\omega_i/\omega_j$ is very small, less than 1%, and can be treated as equal. In other words, $r_j^2\omega_j=r_i^2\omega_i$.

Table 1:$r_j^2/r_i^2$ and $\omega_i/\omega_j$ obtained from the observatory data

| Time | $\mu_i$ | $\mu_j$ | $\alpha_i$ | $\alpha_j$ | $\omega_i$ | $\omega_j$ | $r_j^2/r_i^2$ | $\omega_i/\omega_j$ |
|---|---|---|---|---|---|---|---|---|
| 9, 13, 1942 | | 7.5° | | 4.6° | | 0.974 | 1.005 | 0.995 |
| 7, 31, 1944 | | 33.8° | | 20.2° | | 0.956 | 1.026 | 1.014 |
| 6, 18, 1946 | | 62.4° | | 33.2° | | 0.954 | 1.016 | 1.016 |
| 5, 5, 1948 | 100.2° | | 37.1° | | 0.969 | | | |
| 2, 8, 1952 | | 100.1° | | 36.2° | | 1.012 | 0.959 | 0.958 |
| 12, 26, 1953 | | 60.6° | | 31.5° | | 1.019 | 0.958 | 0.951 |
| 11, 13, 1955 | | 30.3° | | 17.7° | | 1.007 | 0.967 | 0.962 |

All of the data shown in Table 1 are based upon the positions of the Earth and its corresponding angular speed at the opposition of Mars on March 25, 1950, and 1 to 4 Martian years before and after, respectively. Similarly, these procedures can be continued for up to 30 more positions of the Earth, $E_k$, and its corresponding angular speed, $\omega_k$, by including the data from 5 to 20 Martian years before and after the aforesaid opposition (Table 2). The relationship $r_j^2\omega_j=r_i^2\omega_i$ still holds.

Table 2:The values of $\omega_k$, $r_k^2/r_i^2$ and $\omega_i/\omega_k$ obtained on 30 different dates

| Dates | $\omega_i$ | $\omega_k$ | $r_k^2/r_i^2$ | $\omega_i/\omega_k$ | Dates | $\omega_i$ | $\omega_k$ | $r_k^2/r_i^2$ | $\omega_i/\omega_k$ |
|---|---|---|---|---|---|---|---|---|---|
| 5, 5, 1948 | 0.969 | | | | 5, 5, 1948 | 0.969 | | | |
| 10,26, 1940 | | 0.998 | 0.959 | 0.971 | 9,30, 1957 | | 0.983 | 0.982 | 0.986 |
| 12, 9, 1938 | | 1.016 | 0.951 | 0.954 | 8,18, 1959 | | 0.961 | 1.008 | 1.008 |
| 1,21, 1937 | | 1.017 | 0.953 | 0.952 | 7, 5, 1961 | | 0.953 | 1.016 | 1.016 |
| 3, 6, 1935 | | 1.001 | 0.968 | 0.968 | 5,23, 1963 | | 0.962 | 1.005 | 1.008 |
| 4,18, 1933 | | 0.977 | 0.991 | 0.992 | 4, 9, 1965 | | 0.982 | 0.988 | 0.987 |
| 6, 1, 1931 | | 0.958 | 1.008 | 1.012 | 2,25, 1967 | | 1.005 | 0.958 | 0.964 |
| 7,14, 1929 | | 0.954 | 1.019 | 1.016 | 1,12, 1969 | | 1.019 | 0.950 | 0.951 |
| 8,27, 1927 | | 0.966 | 1.011 | 1.003 | 11,29, 1970 | | 1.014 | 0.953 | 0.956 |
| 10, 9, 1925 | | 0.988 | 0.993 | 0.980 | 10,16, 1972 | | 0.992 | 0.974 | 0.976 |
| 11,22, 1923 | | 1.010 | 0.953 | 0.959 | 9, 3, 1974 | | 0.969 | 0.999 | 1.000 |
| 1, 4, 1922 | | 1.019 | 0.951 | 0.951 | 7,21, 1976 | | 0.955 | 1.012 | 1.015 |
| 2,17, 1920 | | 1.009 | 0.957 | 0.960 | 6, 8, 1978 | | 0.957 | 1.010 | 1.013 |
| 4, 1, 1918 | | 0.986 | 0.980 | 0.983 | 4,25, 1980 | | 0.973 | 0.991 | 0.996 |
| 5,14, 1916 | | 0.964 | 1.005 | 1.005 | 3,13, 1982 | | 0.997 | 0.974 | 0.972 |
| 6,27, 1914 | | 0.954 | 1.017 | 1.016 | 1,29, 1984 | | 1.016 | 0.951 | 0.954 |

Because the periods of the Martian year, 687 days, and that of the Earth year, 365 days, are mutually prime, the positions of the Earth recorded every Martian year before and after March 23, 1950 will cover almost every position along the orbit of the Earth around the Sun. The positions selected by the method shown above are so dense as to approach generality. Due to this ergodicity, the law of equal areas may be established: $r^2\omega$=constant.

Certainly, other oppositions of Mars can be examined, for example, Feb. 10, 1916, and Feb. 12, 1995. The difference between these 2 oppositions is only 1.8°. The procedures may be repeated as in Table 2, and the relation of equal areas may be verified: $r^2\omega$=constant. In retrospect, this method for establishing the law of equal areas is, on the one hand, to apply the information on the oppositions of Mars as well as the period of Mars, and, on the other hand, to use the mutual prime property between the 687- and 365-day periods of Mars and Earth, respectively, to guarantee ergodic distributions of the selected positions of the Earth.

It is very hard to directly measure the distances from the Earth or other planets to the Sun, or the ratio of the distances at 2 different positions. Nevertheless, it is relatively easy to observe the angles swept by the Earth. The establishment of the law of equal areas helps us, through measuring the angular speed of the planets, to derive the ratio of distances from the planets to the Sun at different times. This is the implicit meaning of the law of equal areas, which can be used effectively to determine distances from the planets to the Sun.

## 3. The Law of Ellipses for the Earth

After constructing the law of equal areas for the Earth, the next task is to establish the law of elliptical orbits for the Earth. Since the motion of the Earth around the Sun is regularly periodic, the distance function $r(\theta)$ from the Earth to the Sun, or the reciprocal of $r(\theta)$ can also be expressed as a periodic function of $\theta$. Namely, it can be expressed as an infinite series of sines and cosines with different multiple angles as follows [3, 4]:

$$\frac{1}{r} = a_0 + \sum_n a_n \cos n\theta + \sum_n b_n \sin n\theta \qquad (n=1,2,3\ldots)$$

In the ideal case, this function can be approximated by a single period of the trigonometric functions, i.e.,

$$\frac{1}{r} = a_0 + a_1 \cos\theta + b_1 \sin\theta \qquad (2)$$

By applying the law of equal areas as shown in eq. (1), or

$$\frac{1}{r} = c\sqrt{\omega}$$

where $c$ is a proportional constant, the periodic function of the reciprocal of the distance may be

expressed as follows:

$$\sqrt{\omega} = c_o + c_1 \cos\theta + c_2 \sin\theta$$

By doing this, the observable angular speeds $\omega$ may be used to replace the unobservable distances $r$. In order to find the 3 unknown coefficients $c_0$, $c_1$ and $c_2$ as shown in the above equation, one has to choose 3 sets of data to set up simultaneous linear equations with 3 unknowns. One may randomly select 3 sets of data on April 23, 1998, July 31, 1998, and Oct. 2, 1998, to form

$$\sqrt{\omega_1} = c_o + c_1 \cos\theta_1 + c_2 \sin\theta_1$$

$$\sqrt{\omega_2} = c_o + c_1 \cos\theta_2 + c_2 \sin\theta_2$$

$$\sqrt{\omega_3} = c_o + c_1 \cos\theta_3 + c_2 \sin\theta_3$$

where the values of $\theta_1$, $\theta_2$ and $\theta_3$ are the inclined angles between the line connecting the Earth to the Sun at 3 different dates and the x-axis, which is set along the line connecting the Earth to the Sun on Jan. 27, 1998. Hence $\theta_1 = 213.2° - 127.4° = 85.8°$, as shown in Table 3.

Table 3: The reference data for finding the periodic function of the reciprocal of the distance

| Time | Longitude | $\theta_i$ | $\omega_i$ |
|---|---|---|---|
| 1, 27, 1998 | 127.4° | | |
| 4, 23, 1998 | 213.2° | 85.8° | 0.975 |
| 7, 3, 1998 | 308.1° | 180.7° | 0.956 |
| 10, 2, 1998 | 9.1° | -118.3° | 0.984 |

From the values of $\theta_1$, $\theta_2$, $\theta_3$, $\omega_1$, $\omega_2$, and $\omega_3$ in Table 3, the solutions to the above simultaneous linear equations in 3 unknowns can be found. Their solutions are as follows:

$$c_0 = 0.993 \qquad c_1 = 0.015 \qquad c_2 = -0.007.$$

Hence, the periodic function for the square root of angular speed at 3 distinct positions of the Earth, or the reciprocal of the distances from 3 different positions of the Earth to the Sun is as follows:

$$\sqrt{\omega} = 0.993 + 0.015 \cos\theta - 0.007 \sin\theta \qquad (3)$$

Furthermore, 7 different positions of the Earth are randomly selected for 7 different dates, and set $d = 0.993 + 0.015\cos\theta - 0.007\sin\theta$. By comparing the difference of $\sqrt{\omega}$ and $d$ from the corresponding angular speed $\omega$, whether the selected position of the Earth satisfies the periodic function of the distance can be verified, as shown in Table 4. This table shows that $\sqrt{\omega}$ is identical to $d$, and this supports the validity of the periodic function in eq. (3).

In fact, the periodic function shown in equations (2) and (3) is exactly the same equation as that for the path of an elliptical orbit. If one takes the line connecting the 2 foci to be the x-axis, then the path of elliptical orbit can be expressed as follows [5, 6, 7]:

Table 4：Verifying the effectiveness for the periodic function of the distance at 7 different positions

| Time | Longitudes | $\theta$ | $\omega$ | $\sqrt{\omega}-d$ |
|---|---|---|---|---|
| 2,12, 1998 | 143.6° | 16.2° | 1.011 | 0.000 |
| 3,17, 1998 | 176.7° | 49.3° | 0.995 | 0.000 |
| 5, 6, 1998 | 225.8° | 98.4° | 0.968 | 0.000 |
| 6, 22, 1998 | 270.9° | 143.5° | 0.954 | 0.000 |
| 8,18, 1998 | 325.4° | 198.0° | 0.962 | 0.000 |
| 11,17, 1998 | 55.0° | -72.4° | 1.009 | 0.000 |
| 12,29, 1998 | 97.6° | -29.8° | 1.019 | 0.000 |

$$r = \frac{a(1-e^2)}{1+e\cos\theta} \quad \text{or} \quad \frac{1}{r} = B + A\cos\theta \quad ,$$

where $e$ is eccentricity, $a$ is semi-major axis, $B=1/a(1-e^2)$ and $A=e/a(1-e^2)$. If the x-axis is not the line connecting two foci, then the equation for the ellipse is as follows:

$$\frac{1}{r} = B + A\cos(\theta - \theta_0) = B + d_1\cos\theta + d_2\sin\theta \tag{4}$$

where $\theta_0$ is the angle between the line connecting the perihelion to the origin and the x-axis, $d_1^2 + d_2^2 = A^2$, and $\sqrt{d_1^2 + d_2^2}/B = A/B = e$. Hence, the periodic function of the reciprocal of the distance shown in eq. (2) or (3) has the form for the equation of ellipse, and the ratio of $\sqrt{c_1^2 + c_2^2}$ and $c_0$ is as follows:

$$\frac{\sqrt{c_1^2 + c_2^2}}{c_0} = \frac{\sqrt{(-0.007)^2 + 0.015^2}}{0.993} = 0.017$$

which is exactly the same as the accepted value 0.017 for the eccentricity of the Earth's orbit. These can obviously show the equivalence of the periodicity of the reciprocal of the distance of the Earth to the equation of an ellipse, and verify firmly that the orbit of the Earth is exactly elliptical.

## 4. Conclusions

Applying the specific properties of the oppositions of Mars and the Martian year, the position of Mars relative to the Sun can be fixed in the celestial sphere. By means of the geometric relationship formed by these 2 fixed points and the motions of the Earth, one may overcome the obstacles due to the unobservable changes in the distances between the Earth and the Sun, and expresses them in terms of relatively easily measurable angles. Through these, the laws of equal areas and elliptical orbits for the Earth can be concisely established, which should confer a very deep and sincere admiration for Johannes Kepler and the insights he contributed some 400 years ago.